\documentclass{article}

\usepackage{arxiv}

\usepackage[utf8]{inputenc} 
\usepackage[T1]{fontenc}    
\usepackage{hyperref}       
\usepackage{url}            
\usepackage{booktabs}       
\usepackage{amsfonts}       
\usepackage{nicefrac}       
\usepackage{microtype}      
\usepackage{lipsum}
\usepackage{graphicx}
\usepackage{array}
\usepackage{tabularx}
\usepackage{xcolor}
\usepackage{amsmath}
\usepackage{graphicx}
\usepackage{mathtools}
\usepackage{stmaryrd}
\usepackage[ruled,vlined]{algorithm2e}
\usepackage{tcolorbox}
\usepackage{wrapfig}
\usepackage{tikz}
\usepackage{graphicx}
\usepackage{svg}
\usepackage{amssymb}
\usepackage{mathabx}
\usepackage{hyperref}
\usepackage{moresize}
\usepackage[utf8]{inputenc}

\usepackage{lipsum}
\usepackage{subcaption}
\usepackage{scalerel}
\hypersetup{
    colorlinks=true,
    linkcolor=black,
    urlcolor=blue,
    pdftitle={},
    pdfpagemode=FullScreen,
    }
\usepackage{cite}

\makeatletter
\newcommand{\removelatexerror}{\let\@latex@error\@gobble}
\makeatother

\usetikzlibrary{calc, fit, arrows}
\newtheorem{theorem}{Theorem}[section]
\newtheorem{assumption}[theorem]{Assumption}

\newtheorem{remark}[theorem]{Remark}
\newtheorem{problem}[theorem]{Problem}
\newtheorem{proof}[theorem]{Proof}
\title{Quantification of Sim2Real Gap via Neural Simulation Gap Function}

\author{
 P Sangeerth \\
 Centre for Cyber-Physical Systems\\
  Indian Institute of Scince\\
  Bengaluru,India  \\
  \texttt{sangeerthp@iisc.ac.in} \\
   \And
Pushpak Jagtap\\
Centre for Cyber-Physical Systems\\
  Indian Institute of Scince\\
  Bengaluru,India\\
  \texttt{pushpak@iisc.ac.in} \\
}
 
\begin{document}

\maketitle
\pagestyle{empty}

\begin{abstract}

In this paper, we introduce the notion of neural simulation gap functions, which formally quantifies the gap between the mathematical model and the model in the high-fidelity simulator, which closely resembles reality. Many times, a controller designed for a mathematical model does not work in reality because of the unmodelled gap between the two systems. With the help of this simulation gap function, one can use existing model-based tools to design controllers for the mathematical system and formally guarantee a decent transition from the simulation to the real world. Although in this work, we have quantified this gap using a neural network, which is trained using a finite number of data points, we give formal guarantees on the simulation gap function for the entire state space including the unseen data points. We collect data from high-fidelity simulators leveraging recent advancements in Real-to-Sim transfer to ensure close alignment with reality. We demonstrate our results through two case studies - a Mecanum bot and a Pendulum.   

\end{abstract}


\section{Introduction}
Plenty of literature is available for systems whose precise mathematical models are known. However, in real-world scenarios, getting the exact mathematical model of the system is not always possible. Hence, the controllers developed using the approximate mathematical model of the real world do not necessarily need to work on the original system. This introduced the trend of designing data-driven control algorithms that do not use the actual model of the system but demand vast data of the states, input, and output of the system to design a controller to satisfy the desired specifications. \cite{nelles2020nonlinear} contains several ways of identifying the system from the input and output data without guarantees.

Several papers, including \cite{avzman2011dynamical,gorantla2022funnel,jagtap2020control,beckers2019stable} use the Gaussian process (GP) where the system dynamics is unknown, and estimated dynamics is used to design controllers with probabilistic guarantees by collecting data from the system. Data-driven techniques similar to the ones mentioned in \cite{10015033,10266799,nejati2023formal} for control Lyapunov functions, control barrier functions, and robust control techniques.


However, collecting real-world data is challenging due to the significant time and cost involved. The advent of simulators addressed the problem of time by parallelizing the data collection process.The advancements in sensor technology have improved simulation software, reducing the gap between real-world and simulated environments \cite{chen2022real2sim,9363564}. Advanced simulators like ADAMS-Simulink \cite{brezina2011using}, Unreal physics engine \cite{unrealengine}, Metamoto \cite{samak2021autodrive}, NVIDIA's Drive Constellation\footnotemark[3]{}, CarMaker\footnotemark[4]{}, and OpenDRIVE\footnotemark[5]{} enhance simulation fidelity using advanced sensor models, physics engines, and uncertainty models, providing valuable data for developing controllers with data-driven approaches.
\footnotetext[3]{https://resources.nvidia.com/en-us-auto-constellation/drive-constellation}
\footnotetext[4]{https://ipg-automotive.com/en/products-solutions/software/carmaker/}
\footnotetext[5]{https://www.asam.net/standards/detail/opendrive/}

Several studies, such as \cite{DBLP:journals/corr/TobinFRSZA17,10260398,akella2023safety,sangeerth2024towards}, concentrate on bridging the gap between simulation environments and reality within the Sim2Real context. In \cite{akella2023safety}, the author quantifies the Sim2Real gap as a numerical value with a probabilistic guarantee. {In \cite{sangeerth2024towards}, the authors quantified the Sim2Real gap as a function dependent on both state and input with $100\%$ guarantees. However, their approach assumes that this function can be expressed as a linear combination of predefined basis functions, with weight parameters determined through an optimization process similar to regression. The limitation of this method is that if the chosen structure of the simulation gap function is incorrect, the results may become overly conservative. To address this issue, we propose a more flexible approach by modeling the simulation gap function using a neural network, which can better capture complex, nonlinear relationships without requiring a predefined structure.} 
Once this simulation gap is determined, we leverage existing model-based control techniques to synthesize a controller that operates seamlessly with the high-fidelity simulator model.  We demonstrate this claim with two physical case studies: a Mecanum bot (simulator model in Gazebo) and a Pendulum (simulator model in Pybullet). 

\section{System Description and Problem Definition}
\textbf{Notations.} We denote sets of non-negative integers, positive real numbers, and non-negative real numbers, respectively, by $\mathbb{N}:=\{0,1,2,3,\dots\}$, $\mathbb{R}^+$, and $\mathbb{R}^+_0$. The absolute value of $x \in \mathbb{R}$ is denoted by $|x|$. The Euclidean norm of $x \in \mathbb{R}^n$ is represented by $\lVert x\rVert$. The symbol ${\mathbb{R}^n}$ is used to denote an $\textit{n}$-dimensional Euclidean space. A column vector $x \in \mathbb{R}^n$ is denoted by $x\hspace{-0.2em}=\hspace{-0.2em}[x_1;x_2;\cdots;x_n]$. For a matrix $A \in \mathbb{R}^{m \times n}$, $A^\top$ denotes its transpose. The set of diagonal matrices of dimension $n$ is denoted by $\mathcal{D}^n$ and $\mathcal{D}_{\geq 0}^n$ is the set of diagonal matrices with non-negative entries. 
\subsection{System Description}
\label{system description subsection}
In this paper, we consider a nominal mathematical model for the system represented by a discrete-time system, derived through the discretization of conventional continuous-time models \cite{rabbath2013discrete}, with a sampling time $\tau\in\mathbb{R}^+$, expressed as 
\begin{equation}
\label{disc_mathematical_model}
    {\Sigma}\!: x(k+1) = {f}_{\tau}(x(k),u(k)), \quad k \in \mathbb{N},
\end{equation}
where $x =[x_1; x_2; \cdots ; x_n] \in X \subset \mathbb{R}^n$ is the state vector  in a bounded state set $X$, $u=[u_1;u_2;\cdots; u_m] \in U\subset \mathbb{R}^m$ is the input vector in a bounded input set $U$, and ${f}_{\tau}:\textit{X} \times \textit{U} \rightarrow \textit{X}$ is the known transition map which is assumed to be locally Lipschitz continuous to guarantee the uniqueness and existence of the solution \cite{Khalil:1173048}.

Controllers designed for exact mathematical models often underperform in practice. However, advances in sensor technology now provide accurate real-world state measurements, enhancing our understanding of real systems. This leads to the development of high-fidelity simulators like Gazebo \cite{1389727}, CARLA \cite{dosovitskiy2017carla}, Webots \cite{Webots}, and Pybullet \cite{coumans2021} which closely replicate real systems using sensor models, physics engines, and uncertainty models. We represent the real-world control system evolution in these simulators by an unknown discretized map with the same discretization parameter $\tau$ as  
\begin{equation}\label{simulator_model}
    \hat{\Sigma}\!: x(k+1) =\hat{f}_{\tau}(x(k),u(k)), \quad k \in \mathbb{N},
\end{equation}
where $\hat{f}_{\tau}:{X} \times {U} \rightarrow \textit{X}$ represents the transition map of the high-fidelity simulator model, which is unknown. For brevity, the $f_{\tau}$ and $\hat{f}_{\tau}$ will be represented as $f$ and $\hat{f}$ in the further discussions.

\subsection{Problem Formulation}
\label{problem formulation section}
Now, we outline the problem considered in this work. 
 \begin{tcolorbox}[width=\linewidth]
\begin{problem}
    \label{Problem-1}
    Given the nominal mathematical model ${\Sigma}$ and the high-fidelity simulator $\hat{\Sigma}$, we aim 
    \begin{enumerate}
        \item[(i)] to formally quantify a simulation gap between the mathematical model and the high-fidelity simulator model, represented by a function $\gamma(x,u)\!:X \times U \rightarrow {\mathbb{R}^+_0}^n $ such that for all $x \in X,$ and $u \in U $,
        {\begin{equation}
        \label{problem_statement_equation}
            |\hat{f}_i(x,u)-{f}_i(x,u)| \leq \gamma_i(x,u)
        \end{equation}} for all $i \in \{1,2,\ldots,n\}$ and
        \item[(ii)] to design a controller using the nominal mathematical model and the neural simulation gap $\gamma(x,u):=[\gamma_1(x,u);\gamma_2(x,u);\ldots;\gamma_n(x,u)]$ and to enforce desired specifications in the high-fidelity simulator ${\hat\Sigma}$.
    \end{enumerate}
\end{problem}
\end{tcolorbox}

\section{Neural Simulation Gap Function}\label{Neural Simulation gap function}
An unpreventable gap exists between the mathematical model and the high-fidelity simulators. This section focuses on measuring this gap using a neural simulation gap function.

{To compute the simulation gap function, we first reformulate the conditions mentioned in \eqref{problem_statement_equation} as a robust optimization problem (ROP) as follows:} 
\begin{equation}
\label{eq: ROP_for_one_state}
\begin{aligned}
\min_{\gamma_i \in \mathcal{H},\eta_i \in \mathbb{R}}
&\eta_i  \\
\text{s.t.}\quad\quad &\gamma_i(x,u) \leq \eta_i, \quad \forall x\!\in\! X, \forall u \!\in\! U,\\
& |\hat{f_i}(x,u)-{f_i}(x,u)| \!-\! \gamma_i(x,u) \!\leq\! 0,\quad \forall x\!\in\! X, \forall u \!\in\! U,\\
\end{aligned}
\end{equation}
where $i \in I$, $I:=\{1,2,\dots,n\}$ and $\mathcal{H}:=\{g \,\big|\, g\!:X \times U \rightarrow \mathbb{R}_0^+\}$ is a functional space containing the set of all possible functions. For brevity, whenever we use the subscript $i$, the full set $I$ is referred to unless specified otherwise.
One can readily observe considerable challenges in solving the ROP in \eqref{eq: ROP_for_one_state}. Firstly, the ROP in \eqref{eq: ROP_for_one_state} has infinite constraints due to continuous state and input spaces. Secondly, the function map $\hat{f_i}(x,u)$ in \eqref{eq: ROP_for_one_state} and the structure of $\gamma_i(x,u)$ are unknown. To address these challenges, we start with fixing the structure of $\gamma_i(x,u)$. {As done in \cite{sangeerth2024towards}, one can fix the structure of} $\gamma_i(x,u)=\sum_{l=1}^{z_i} w^{(l)}_ip^{(l)}_i(x,u)$, a parametric form linear in the decision variable $w_i= [w^{(1)}_i;\cdots;w^{(z_i)}_i] \in \mathbb{R}^{z_i}$ with user-defined basis functions $p^{(l)}_i(x,u)$. But this method is prone to bad approximation because of improper choices of basis functions. We address this issue by fixing $\gamma_i(x,u)$ as a neural network $\gamma_i^{w}(x,u)$ (universal function approximator) for all $i \in I$ parameterized by weights represented by the variable $w$.

Given systems $\Sigma$ and $\hat{\Sigma}$, $\gamma_i^{w}(x,u)$ represent the neural simulation gap function for all $i \in I$ which is parameterized by weights $w$. $\gamma_i^{w}(x,u)$ consists of an input layer with $n+m$ (i.e., system dimension and input dimension) neurons and an output layer with one neuron due to the scalar output of the simulation gap function for each state. Moreover, the number of hidden layers (depth) is denoted by $l_b$. Similarly, for every hidden layer $1 \leq j \leq l_b$, the number of neurons in that layer is denoted by $h^j_b$. The activation function in all the layers can chosen to be any standard activation functions like arctangent, softmax, ReLU. The resulting neural network function is obtained by recursively applying the activation functions at every layer as
\begin{align}
    \mathsf{x}^0&=(x,u) \in X \times U,\nonumber\\
    \mathsf{x}^{j+1}&=\sigma^{(j)}(w^{(j)}\mathsf{x}^{(j)}+b^{(j)}), \forall j \in \{1,2,\ldots,l_b-1\},\nonumber\\
    \gamma_i^{(w)}(x,u)&=\sigma^{(j)}(w^{(l_b)}\mathsf{x}^{(l_b)}+b^{(l_b)}),
\end{align}
where $\sigma^{(j)}(.)$ and $\sigma^{(l_b)}(.)$ are the activation functions applied element-wise on the layers for all $j \in \{1,2,\ldots,l_b-1\}$ and in the last layer $l_b$ respectively. 

In order to train the above neural network, we generate data as described further. We construct the cover sets by appropriately selecting $N$ samples from state set $X$ and $M$ samples from the input set $U$. We initially gather a set of $N$ sampled data points within $X$ by considering balls $X_r$ around each sample $x_r$, $r\in\{1,\ldots,N\}$ with radius $\epsilon_x$ such that $X \subseteq \bigcup_{r=1}^NX_r$ and 
\begin{equation}
\label{max_epsilon_x}
    \Vert x-x_r \Vert \leq \epsilon_x, \quad \forall x \in X_r.
\end{equation}
Repeating the same procedure for the input space $U$ by collecting a finite set of $M$ data points by constructing balls $U_s$ around each sample $u_s$, $s \in \{1,2,\ldots,M\}$ with radius $\epsilon_u$ such that $U \subseteq \bigcup^{M}_{s=1}U_s$.
\begin{equation}
\label{max_epsilon_u}
    \Vert u-u_s \Vert \leq \epsilon_u, \quad \forall u \in U_s.
\end{equation}
Using these $N$-representative points $x_r \in X_r$ as initial conditions and the $M$ input representative points $u_s \in U_s$ as input, we collect $N \times M$ data from the mathematical model and the high-fidelity simulator for the next sampling instance, for all $i \in \{1,2,\ldots,n\}$, which is represented as 
 {\small \begin{align}
 \label{data_eqn}
    \big\{\hspace{-0.1em}(x_r,u_s,\hat{f_i}(x_r,u_s),{f_i}(x_r,u_s))\big|r\hspace{-0.1em}\in\hspace{-0.1em}\{1,..,N\}\hspace{-0.1em},\hspace{-0.1em}s \hspace{-0.1em}\in\hspace{-0.1em} \{1,..,M\}\}.
\end{align}}

The ROP described by \eqref{eq: ROP_for_one_state} for all $i \in I$, is now reformulated as the following scenario convex program (SCP) based on data:
\begin{align}
\min_{\eta_i \in \mathbb{R}}\  &\ \ \eta_i  \nonumber\\
\text{s.t.}\quad & \gamma_i^{(w)}(x_r,u_s) \leq \eta_i, \quad \forall r\in\{1,\ldots,N\}, \forall s \!\in\! \{1,\ldots,M\}, \nonumber\\
& |\hat{f_i}(x_r,u_s)-{f_i}(x_r,u_s)|-\gamma_i^{(w)}(x_r,u_s)\leq 0, \nonumber\\&\forall r\in\{1,\ldots,N\}, \forall s \in \{1,\ldots,M\}.\label{eq: SOP_for_one_state}
\end{align}

One can notice that casting \eqref{eq: ROP_for_one_state} as the above-mentioned SCP in \eqref{eq: SOP_for_one_state} solves the aforementioned challenges. However, training neural networks in this manner will not guarantee that the simulation gap function obtained from \eqref{eq: SOP_for_one_state} will satisfy \eqref{eq: ROP_for_one_state} for unseen data points. To provide guarantees for all the data points of the state space $X$ and input space $U$, we need Lipschitz continuity of the neural network along with the satisfaction of \eqref{eq: SOP_for_one_state}. To render the optimization program mentioned in \eqref{eq: SOP_for_one_state} convex, we first assume that the neural network candidates correspond to the simulation gap function are given to us. These details are explained elaborately in the next section.

\section{Training Neural Simulation Gap Function}
\label{sec-training of nn}
In this section, we will see how the optimization problem given by equation \eqref{eq: SOP_for_one_state} can be solved along with training the neural simulation gap function.

We now discuss the procedure to train the neural network representing simulation gap function $\gamma_i^{(w)}(x,u)$ between $\Sigma$ and $\hat{\Sigma}$, for all $i \in I$, such that the trained simulation gap function satisfies the conditions mentioned in the constraints of \eqref{eq: SOP_for_one_state}. Note that $\gamma_i^{(w)}(x,u)$ satisfy the Lipschitz continuity assumption since they consist of layers with continuous activation functions.
First, observe that for all $i \in I$, $\gamma_i^{(w)}(x,u)$ is Lipschitz continuous with respect to the concatenated vector $(x,u) \in ({X} \times {U})$, for all $x,y \in X$, for all $u,w \in U$ with a bound $\mathcal{L}^{(i)}_1$ if:
\begin{align}
\label{lipschitz_for_gamma}
    \Vert \gamma_i^{(w)}(x,u) -\gamma_i^{(w)}(y,w) \Vert &\leq \mathcal{L}^{(i)}_1 \Vert \begin{bmatrix}
        x\\
        u
    \end{bmatrix}-\begin{bmatrix}
        y\\
        w
    \end{bmatrix}\Vert,
\end{align}
Then, the Lipschitz constant of $\gamma_i^{(w)}(x,u)$ is bounded by $\mathcal{L}^{(i)}_1$ if
the following matrix inequality of $M^{(i)}(w,\Lambda)$ holds.

\begin{equation}
    \begin{bmatrix}
        \mathcal{L}^{(i)}_1 I_n & -{w^{0}}^\top\Lambda_1 & 0                         & \cdots                                & 0\\
        -\Lambda_1^\top w^{0}   & 2 \Lambda_1            & \cdots                    & 0                                     & 0\\
        0                       & \cdots                 & \ddots                    & -{w^{l_b-1}}^\top                     & 0\\
        \vdots                  &                        & -\Lambda_{l_b}w^{l_b-1}   & 2\Lambda_{l_b}                        & -{w^{l_b}}^\top\\
        0                       & \cdots                 & 0                         & -w^{l_b}                              & I_{h_b^{l_b}}
    \end{bmatrix}\hspace{-0.5em}\geq\hspace{-0.2em}0,
\end{equation}
where $\Lambda=(\Lambda_1,\Lambda_2,\ldots,\Lambda_{l_b})$, $\Lambda_j \in \mathcal{D}^{h_b^j}, j \in \{1,2,\ldots,l_b\}$. We
refer the readers to \cite{fazlyab2019efficient,pauli2022neural,pauli2021training}, for more details on obtaining the above matrix inequality. We now describe the construction of suitable loss functions, inspired by approaches provided in \cite{anand2023formally,DBLP:conf/cdc/TayalZJ0K24}, for the training of $\gamma_i^{(w)}(x,u)$, such that the minimization of these loss functions leads to the satisfaction of conditions required by the SCP according to \eqref{eq: SOP_for_one_state} over the training data sets \eqref{data_eqn}. Consider the following sub-loss functions characterizing conditions
\begin{align}
    L_1^{(i)}(w)=&\max(0,\gamma_i^{(w)}(x,u)-\eta_i), \\
    L_2^{(i)}(w)=&\max(0,|\hat{f_i}(x,u)-{f_i}(x,u)|-\gamma_i^{(w)}(x,u)).
\end{align}
We combine loss functions as follows.
\begin{align}
    \label{net_loss_eqn}
    L^{(i)}(w)=&c_1 \max(0,\gamma_i^{(w)}(x,u)-\eta_i)+\\ \nonumber
    &c_2 \max(0,|\hat{f_i}(x,u)-{f_i}(x,u)|-\gamma_i^{(w)}(x,u)),
\end{align}
where $c_1$ and $c_2$ are some positive constants.
Furthermore, consider an additional loss function, 
\begin{align}
\label{sublossfun}
    L_M^{(i)}(w,\Lambda)=-c\log \det(M^{(i)}(w,\Lambda)),
\end{align}
where $c$ is a positive weight coefficient for the sub-loss functions in \eqref{sublossfun}.

\begin{theorem}
 \label{Theorem 2}
    Consider systems $\Sigma$ and $\hat{\Sigma}$ whose state-space is $X$, and input-space is $U$. If the loss function given by \eqref{net_loss_eqn} satisfies $L^{(i)}=0$ and the additional loss function given by \eqref{sublossfun} satisfies $L_M^{(i)} \leq 0$ for all $i \in I$, then $|\hat{f_i}(x,u)\hspace{-0.2em}-\hspace{-0.2em}{f_i}(x,u)|\leq \gamma_i^{(w*)}(x,u)$ over the training data sets \eqref{data_eqn}, where $\gamma_i^{(w*)}(x,u)$ represents the feasible solution at the optimal value of the SCP. 
\end{theorem}
{\begin{proof}
The proof is very similar to \cite[Theorem 3]{DBLP:conf/cdc/TayalZJ0K24} and is omitted here due to space constraints.
\end{proof}}


{\begin{remark}
\label{bisection_optim}
    The optimization problem is solved using the bisection method, which takes a longer time to converge based on the tolerance set. Exploring alternative methods to solve the optimization problem and reducing the time required for this task needs further investigation as a future work.
\end{remark}}

\section{Simulation gap with Formal Guarantees}
\label{sec-Simulation gap with Formal Guarantees}
In this section, we will formally prove how the neural network trained with a finite dataset will be able to estimate the simulation gap for the unseen data points of the state space.

\begin{assumption}\label{A1}
 Let the function $|\hat{f_i}(x,u)-{f_i}(x,u)|$ be Lipschitz continuous with respect to $x$ with Lipschitz constant $\mathcal{L}^{(i)}_{2x}$, for any finite $u \in U$, and be Lipschitz continuous with respect to $u$ with Lipschitz constants $\mathcal{L}^{(i)}_{2u}$, for any finite $x \in X$, for all $i \in I$. 
\end{assumption}
$\gamma_i^{(w)}(x,u)$ is Lipschitz continuous with respect to the concatenated vector $(x,u)\in (X \times U)$ with Lipschitz constant $\mathcal{L}^{(i)}_1$ as given by \eqref{lipschitz_for_gamma} for all $i \in I$. Even though the function $\hat{f_i}(x,u)$ is not known, one can employ the proposed results in \cite{Wood1996EstimationOT} to estimate the Lipschitz constants of the function mentioned in Assumption~\ref{A1}, using a finite number of data collected from the mathematical model and the simulator model. Under Assumption~\ref{A1}, we now present the following theorem, as the main result of the work, to compute the actual upper bound for the $|\hat{f_i}(x,u)-{f_i}(x,u)|$. 

\begin{theorem} \label{Theorem 1}
Let Assumption \ref{A1} hold. Let the optimal value of SCP in \eqref{eq: SOP_for_one_state} be $\eta_i^{SCP}$. Let the corresponding neural network for which $\eta_i^{SCP}$ is obtained be denoted as $\gamma_i^{(w*)}(x,u)$. Then for all $i \in I$, the absolute difference between the state evolutions of $\Sigma$ and $\hat\Sigma$ is quantified as:
\begin{equation}
\label{actual_gamma}
    |\hat{f_i}(x,u)-{f_i}(x,u)| \leq \gamma_i^{(w*)}(x,u) + \mathcal{L}^{(i)}, 
\end{equation}
for all $x \in X$, for all $u \in U$, where the constant $\mathcal{L}^{(i)}= \mathcal{L}^{(i)}_1 \sqrt{\epsilon_x^2+\epsilon_u^2 }+\mathcal{L}^{(i)}_{2x} \epsilon_x +\mathcal{L}^{(i)}_{2u} \epsilon_u$ and $\epsilon_x,\epsilon_u$ as in \eqref{max_epsilon_x}, and \eqref{max_epsilon_u} respectively.
\end{theorem}
{\begin{proof}For a given $i\in I$, for all $x\in X$ and for all $u\in U$, we have
\begin{align*}
|\hat{f_i}(x,u)\hspace{-0.2em}
&-\hspace{-0.2em}{f_i}(x,u)|=|\hat{f_i}(x,u)\hspace{-0.2em}-\hspace{-0.2em}{f_i}(x,u)|\hspace{-0.2em}\\&-\hspace{-0.2em}|\hat{f_i}(x_r,u_s)\hspace{-0.2em}-\hspace{-0.2em}{f_i}(x_r,u_s)|+|\hat{f_i}(x_r,u_s)\hspace{-0.2em}-\hspace{-0.2em}{f_i}(x_r,u_s)|,
\end{align*}
where $x_r\in X_r$ such that $\|x-x_r\|\leq\epsilon_x$ and $u_s \in U_s$ such that $\|u-u_s\|\leq\epsilon_u$. Under Assumption \ref{A1} along with $\gamma_i^{w*}(x,u)$ obtained by solving SCP in \eqref{eq: SOP_for_one_state}, and from \eqref{lipschitz_for_gamma} one can obtain the following series of inequalities:
\begin{align*}
&|\hat{f_i}(x,u)-{f_i}(x,u)|\\
&\leq \mathcal{L}^{(i)}_{2x} \Vert x-x_r \Vert +\mathcal{L}^{(i)}_{2u} \Vert u-u_s \Vert +|\hat{f_i}(x_r,u_s)\hspace{-0.5em}-\hspace{-0.5em}{f_i}(x_r,u_s)|,\\
&\leq \mathcal{L}^{(i)}_{2x} \epsilon_x  +\mathcal{L}^{(i)}_{2u} \epsilon_u+\gamma_i^{(w*)}(x_r,u_s),\\
&= \mathcal{L}^{(i)}_{2x} \epsilon_x \hspace{-0.2em} +\hspace{-0.2em} \mathcal{L}^{(i)}_{2u} \epsilon_u\hspace{-0.2em} +\hspace{-0.2em} \gamma_i^{(w*)}(x_r,u_s)\hspace{-0.25em}-\hspace{-0.25em}\gamma_i^{(w*)}(x,u)\hspace{-0.25em}+\hspace{-0.25em}\gamma_i^{(w*)}(x,u),\\
&\leq \mathcal{L}^{(i)}_{2x} \epsilon_x\hspace{-0.2em}  +\hspace{-0.2em} \mathcal{L}^{(i)}_{2u} \epsilon_u \hspace{-0.2em} +\hspace{-0.2em} \mathcal{L}^{(i)}_1 \hspace{-0.2em} \sqrt{\Vert x\hspace{-0.2em}-\hspace{-0.2em}x_r \Vert ^2 \hspace{-0.2em}+\hspace{-0.2em}\Vert u\hspace{-0.2em}-\hspace{-0.2em}u_s \Vert ^2 } \hspace{-0.2em}+\hspace{-0.2em}\gamma_i^{(w*)}(x,u),\\
&\leq \mathcal{L}^{(i)}_{2x} \epsilon_x +\mathcal{L}^{(i)}_{2u} \epsilon_u +\mathcal{L}^{(i)}_1 \sqrt{\epsilon_x^2+\epsilon_u^2 } +\gamma_i^{(w*)}(x,u), \nonumber \\
&=\mathcal{L}^{(i)}+\gamma_i^{(w*)}(x,u),
\end{align*}
where $\mathcal{L}^{(i)}=\mathcal{L}^{(i)}_1 \sqrt{\epsilon_x^2+\epsilon_u^2 }+\mathcal{L}^{(i)}_{2x} \epsilon_x +\mathcal{L}^{(i)}_{2u} \epsilon_u $. This concludes the proof.
\end{proof}}
By following equations \eqref{problem_statement_equation} and \eqref{actual_gamma}, we quantify the simulation-gap function for the entire state-space as $\gamma_i(x,u):=\gamma_i^{(w*)}(x,u) + \mathcal{L}^{(i)}$.

\section{Controller Synthesis}
We now design a controller for the high-fidelity simulator system $\hat{\Sigma}$, where the dynamics {of the simulator model} is now represented with the help of nominal mathematical model \eqref{disc_mathematical_model} and the formally quantified neural simulation gap function \eqref{actual_gamma} as
\begin{equation}
    \label{basic_bound_equation}
    x(k+1) \in {f}(x(k),u(k))+[-\gamma(x(k),u(k)),\gamma(x(k),u(k))].
\end{equation}
This system can be perceived as an uncertain system with bounded disturbance, 
a topic extensively covered in existing literature on controller design for uncertain systems. Examples include model predictive control (MPC) \cite{MAYNE2005219}, adaptive control \cite{singh2022adaptive}, and symbolic control \cite{tabuada2009verification}. In our work after calculating the simulation gap function, for the mecanum bot and the pendulum we have designed symbolic controller using SCOTS toolbox \cite{rungger2016scots}. The implementation details of the controller are not provided due to space constraints, and readers are encouraged to refer to \cite{Reissig_2017,rungger2016scots} for more details.

\section{Case Study}

To show the effectiveness of our results, we demonstrate our data-driven approach to a \emph{kinematics model of mecanum bot} and a \emph{pendulum system}. We used a computer with AMD Ryzen 9 5950x, 128 GB RAM to collect data. For the mathematical model, the data was collected from MATLAB for both case studies. In the Mecanum example, the high-fidelity simulator model data was collected from Gazebo, while for the Pendulum example, the high-fidelity simulator model data was collected from PyBullet.

\subsection{Mecanum bot}
\begin{figure}[!ht]
    \centering
    \hspace{-.2em}\includegraphics[scale=0.1]{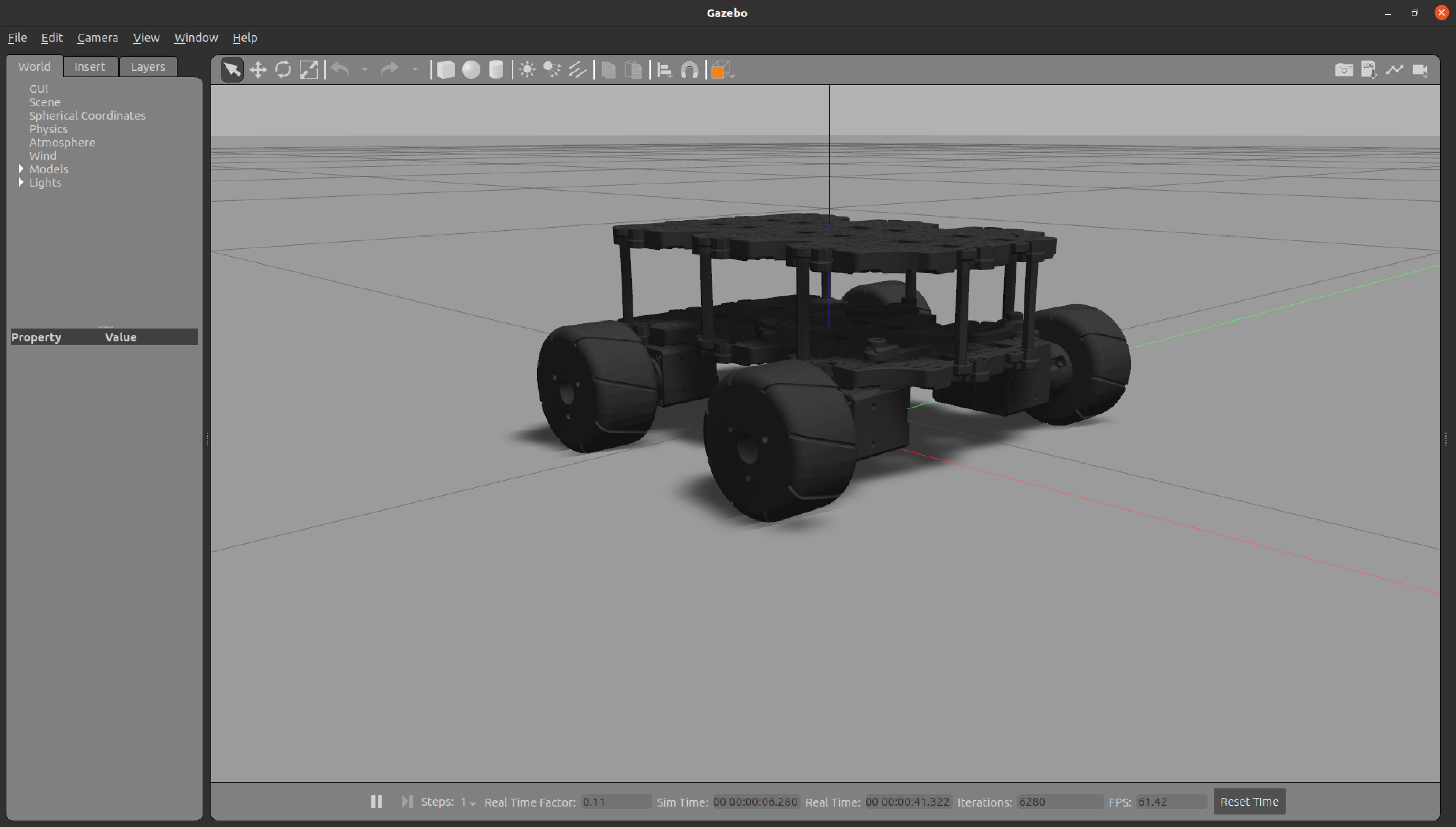}
    \caption{Mecanum model in the Gazebo simulator.}
    \label{fig: Mecanum Gazebo}
\end{figure}
For the first case study, we consider a kinematics model for a differential drive as the following: 
\begin{align*}
    \begin{bmatrix}
    x_1(k+1)\\
    x_2(k+1)\\
\end{bmatrix}=
\begin{bmatrix}
    x_1(k)+\tau u_1(k)\\
    x_2(k)+\tau u_2(k)\\
\end{bmatrix},
\end{align*}
where $x_1$, $x_2$, denotes the position of the mecanum bot in the x and y axes respectively. The parameters $u_1$ and $u_2$ denote the bot's linear velocities in x and y axes respectively. The sampling time is chosen as $0.3s$. The Mecanum bot in a Gazebo simulator is shown in Fig. \ref{fig: Mecanum Gazebo}. The state and finite input sets are, respectively, considered as $X=[0,3]\times[0,3]$ and $U=[-1,1] \times [-1,1]$. The data is collected with the state-space discretization $\epsilon_x=0.01$ input-space discretization $\epsilon_u=0.014$. As mentioned in the remark \ref{bisection_optim}, we fix the hyperparameters of the neural network along with initial value of $\eta_i$ and train the neural network using the bisection method. We select their Lipschitz constants $\mathcal{L}_1^{(1)}$ and $\mathcal{L}_1^{(2)}$ as 10 and 10, respectively, with single hidden layer having 400 neurons for both $\gamma_1(x,u)$, and $\gamma_2(x,u)$. The solution of the optimization problem yields $\eta_1=0.11$ and $\eta_2=0.08$. To compute the simulation gap function $\gamma_i(x,u)$ for the entire state-space with guarantees, we compute the Lipschitz constants of the function $|\hat{f_i}(x,u)-{f_i}(x,u)|$ (refer to assumption-\ref{A1}) using the results of \cite{Wood1996EstimationOT} as $\mathcal{L}_{2x}^{(1)}=1.03,\mathcal{L}_{2x}^{(2)}=1.02,\mathcal{L}_{2u}^{(1)}=1.03,\mathcal{L}_{2u}^{(2)}=1.01$. The total time involved in data collection and solving the optimization problem is approximately $1.5$ days.

\begin{figure}
     \centering
     \begin{subfigure}[b]{0.5\textwidth}
         \centering
         \includegraphics[width=\textwidth]{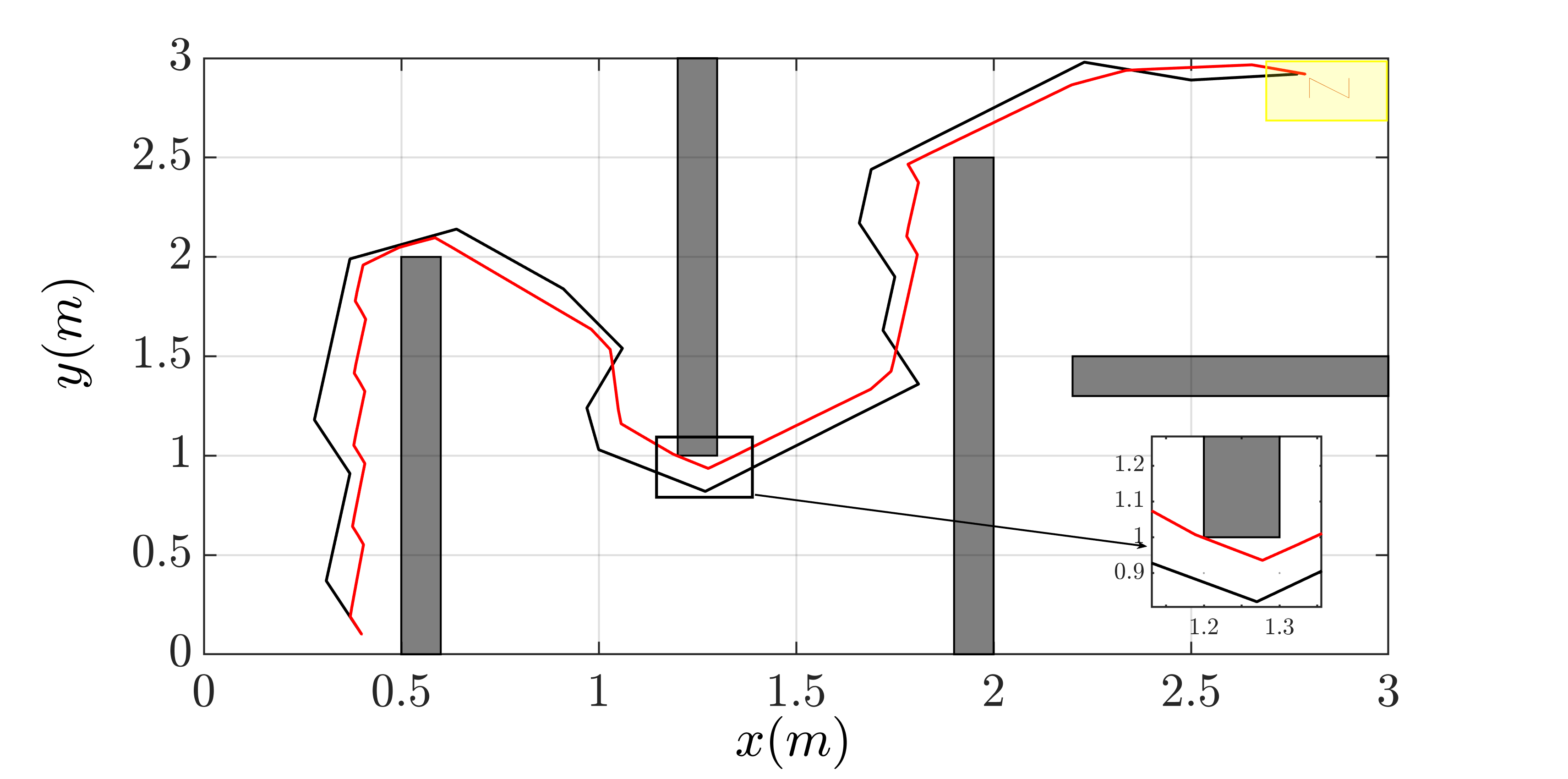}
     \end{subfigure}
     \begin{subfigure}[b]{0.5\textwidth}
         \centering
         \includegraphics[width=\textwidth]{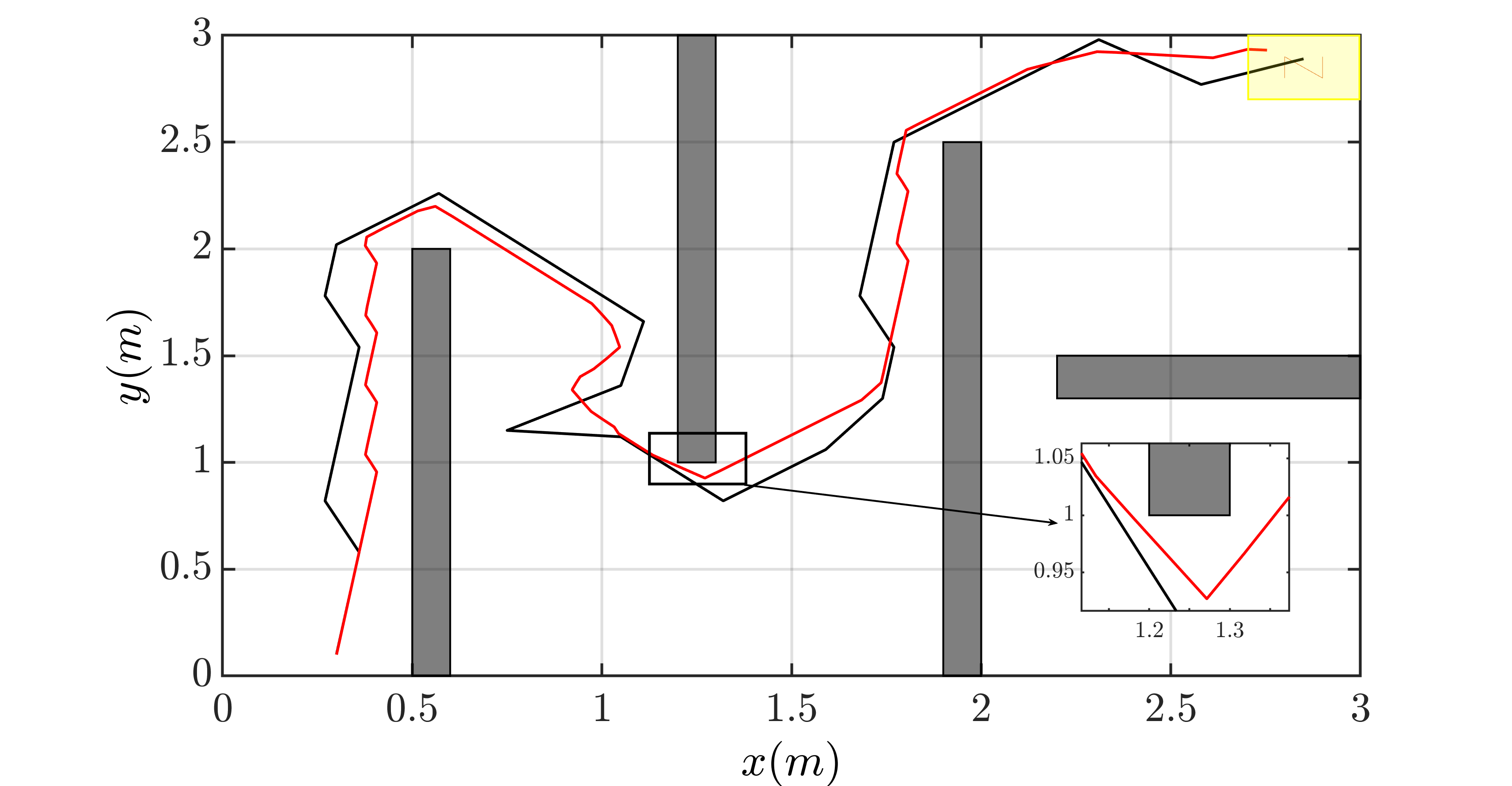}
     \end{subfigure}
        \caption{State trajectories of both the mathematical model (black) and the Gazebo model (red). The black regions represent the obstacles, while the yellow one represents the target. When the controller is synthesized for the reach-while-avoid specification without incorporating $\gamma(x,u)$, the Pybullet model hits the obstacles (top). The underlying specification is satisfied when the controller is synthesized after incorporating $\gamma(x,u)$ (bottom). }
        \label{fig: mecanum plots}
\end{figure}

Using the SCOTS toolbox \cite{rungger2016scots}, we design a symbolic controller for the mathematical model to satisfy the reach-while-avoid specification. However, applying this controller directly to the simulation model leads to failure, causing the robot to hit an obstacle (Fig. \ref{fig: mecanum plots}, top). To address this, we incorporate the simulation-gap function, $\gamma(x,u)$, into the mathematical model’s dynamics and synthesize a new controller. As shown in Fig. \ref{fig: mecanum plots} (bottom), the updated controller successfully meets the specification. This approach enables controller design for an accurate model (Gazebo Simulator) while leveraging a simplified kinematic model and the quantified simulation gap.

\subsection{Pendulum System}
\label{pendulum}
For the second case study, we consider a pendulum system whose mathematical model is given as follows:
\begin{figure}
\vspace{0.5cm}
\centering
  \includegraphics[scale=0.18]{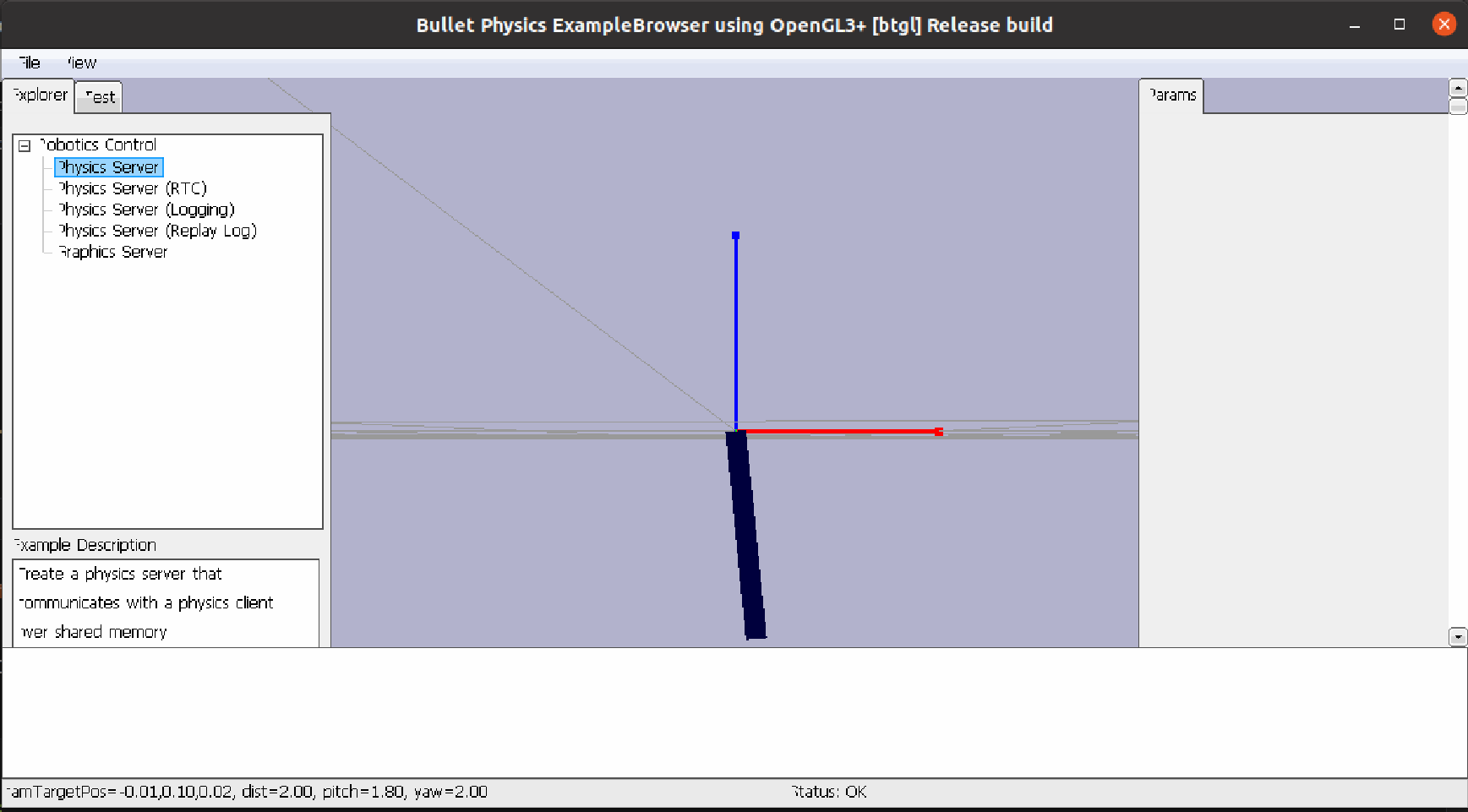}
  \caption{Pendulum model in the Py-Bullet simulator.}
  \label{fig:Pendulum model in Py-Bullet simulator}
\end{figure}
\begin{equation*}
    \begin{bmatrix}
    x_1(k+1)\\
    x_2(k+1)
\end{bmatrix}=
\begin{bmatrix}
    x_1(k)+\tau x_2(k)\\
    -\frac{3g\tau}{2l}\sin{x_1(k)}+x_2(k)+\frac{3\tau u(k)}{ml^2}\\
\end{bmatrix}\!,
\end{equation*}
where $x_1$, $x_2$, and $u$ are the angular position, angular velocity, and torque input, respectively. The parameters $m=1kg$, $g=9.81m/s^2$, and $l=1m$ are, respectively, the pendulum's mass, acceleration due to gravity, and rod length. The pendulum model is simulated in the high-fidelity simulator PyBullet, depicted in Fig. \ref{fig:Pendulum model in Py-Bullet simulator}. The parameter $\tau$ is the sampling time chosen as $0.005s$. The state-space is $X=[-0.2,0.2] \times[-0.25,0.25]$, while the input space is considered $U=[-1,1]$. 
The data is collected with the state-space discretization $\epsilon_x=0.0022$ and input-space discretization $\epsilon_u=0.01$. Following a similar approach to that of the previous case study, we now employ a symbolic control approach to synthesize controllers that enforce an invariance property (\emph{i.e.,} ensuring that the state remains within $[0,0.2]\times[-0.25,0.25]$). These controllers are synthesized using the symbolic controller toolbox SCOTS \cite{rungger2016scots} for both the nominal mathematical model and the model incorporating the simulation gap in \eqref{basic_bound_equation}.  
\begin{figure}[!ht]
    \centering
    \includegraphics[scale=0.16]{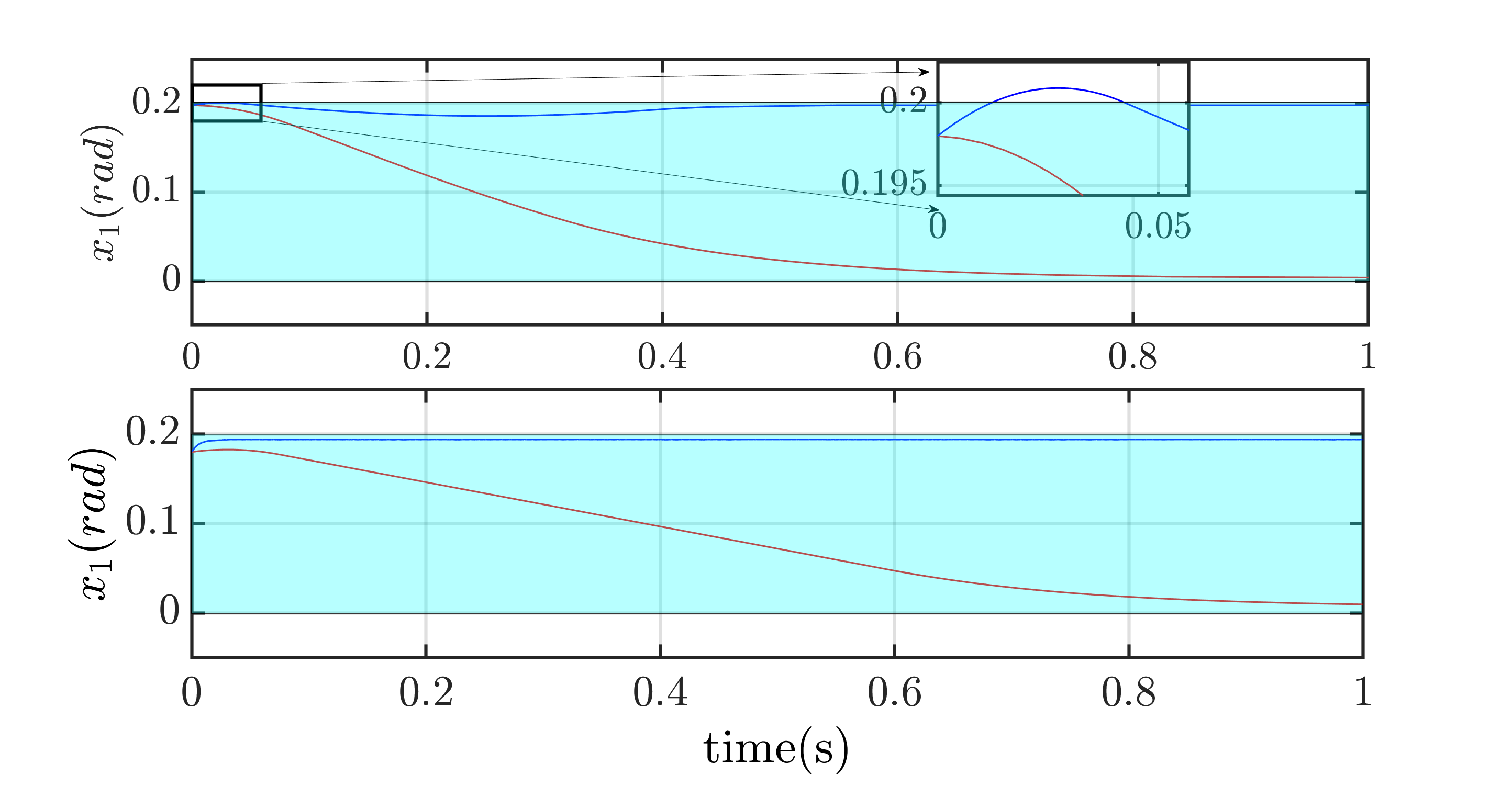}
    \caption{
    The state trajectory $x_1$ for both the mathematical system (red) and the PyBullet model (blue) is shown. The invariance specification is violated in PyBullet when the controller is synthesized without considering $\gamma(x,u)$ (top). The state-space invariance specification is satisfied in PyBullet when the controller is synthesized considering $\gamma(x,u)$ (bottom). The invariance condition on state $x_2$ was satisfied for both the cases without and with $\gamma(x,u)$ and is not showed here for brevity. }
    \label{fig:combined pendulum result}
\end{figure} 
Fig. \ref{fig:combined pendulum result} depicts the trajectories of the mathematical model (red) and the PyBullet model (blue) using controllers designed from the nominal mathematical model without (top) and with (bottom) the quantified simulation gap $\gamma(x,u)$. It is clear that the controller synthesized for the nominal mathematical model results in a violation of the invariance property when used for the PyBullet environment. 
\begin{figure}
    \centering
    \includegraphics[scale=0.16]{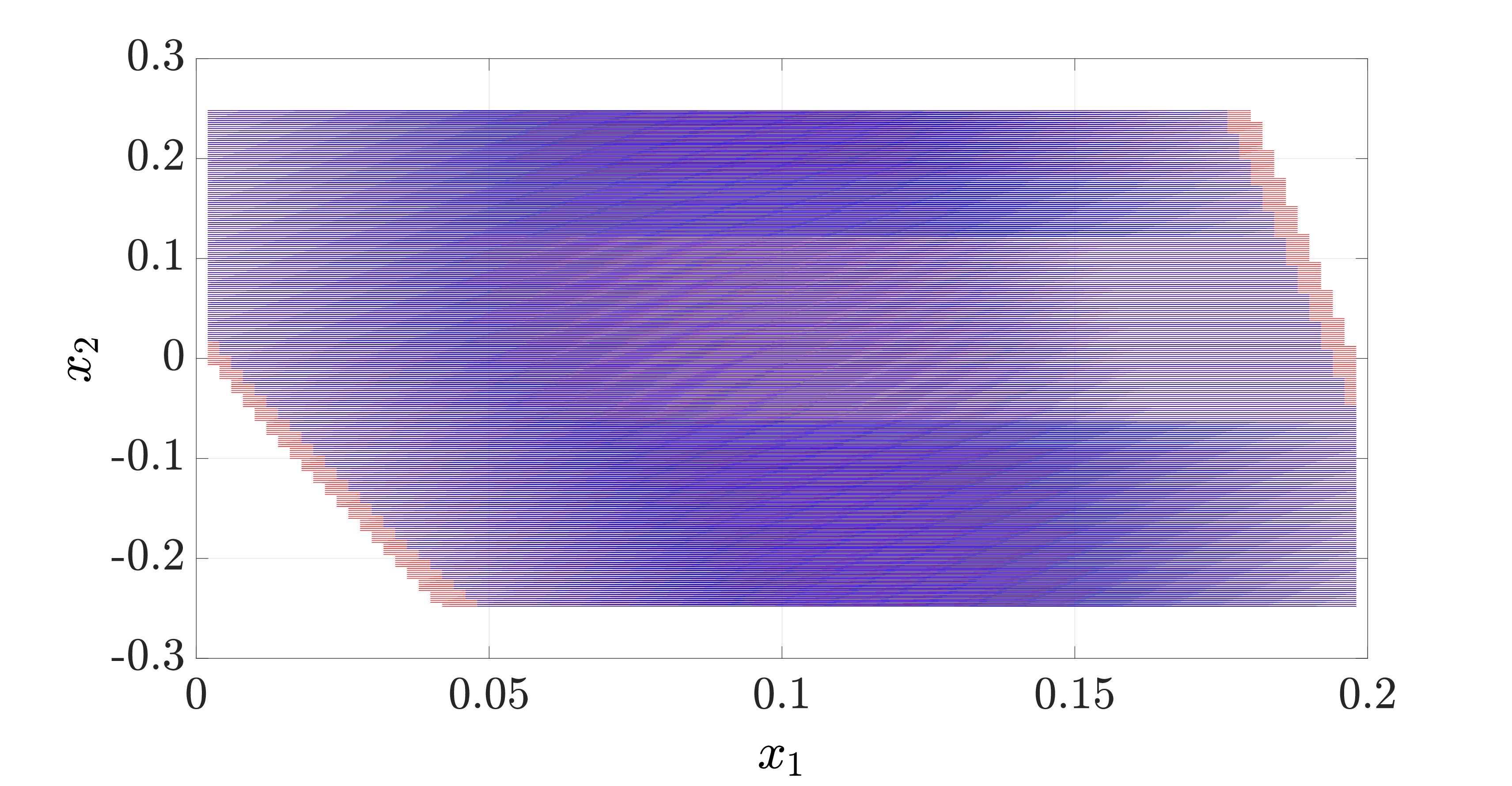}
    \caption{
    The red points represent the \emph{region of invariance} obtained by synthesizing the controller for the mathematical model without $\gamma(x,u)$. The blue points represent the region of invariance obtained by synthesizing after incorporating $\gamma(x,u)$ into the mathematical model as given in in \eqref{basic_bound_equation}.}
    \label{fig:winning domain pendulum}
\end{figure}

\section{Acknowledgements}
We thank Dr. Abolfazl Lavaei for the valuable discussions. 

\section{Conclusion and Future Work}
We present a formal method to quantify the simulation gap function between a mathematical model and a simulator. Our method enables the use of model-based control techniques to meet desired specifications. If the simulator closely resembles reality, incorporating the simulation gap function allows seamless controller deployment in the real world. Our future work will extend this approach to real-world.






\bibliographystyle{ieeetr} 
\bibliography{sources} 

\end{document}